\documentclass[
11pt,tightenlines,%
nofootinbib,
preprintnumbers,%
amsfonts,%
prd]{revtex4}
\usepackage{graphicx}
\newcommand{\be}{\begin{equation}}
\newcommand{\ee}{\end{equation}}
\newcommand{\bea}{\begin{eqnarray}}
\newcommand{\eea}{\end{eqnarray}}

\newcommand{\0}{\over }

\newcommand{\real}{\mathrm{Re}\,}
\newcommand{\imag}{\mathrm{Im}\,}

\newcommand{\Li}{\mathrm{Li}}
\newcommand{\emu}{\varepsilon }

\begin{document}

\preprint{TUW-05-01}

\title{Fermionic dispersion relations in ultradegenerate relativistic
plasmas\\ beyond leading logarithmic order}
\date{\today}
\author{Andreas Gerhold}
\affiliation{Institut f\"ur Theoretische Physik, Technische
Universit\"at Wien, \\Wiedner Hauptstr.~8-10, 
A-1040 Vienna, Austria }
\author{Anton Rebhan}
\affiliation{Institut f\"ur Theoretische Physik, Technische
Universit\"at Wien, \\Wiedner Hauptstr.~8-10, 
A-1040 Vienna, Austria }
\begin{abstract}
We determine the dispersion relations of fermionic quasiparticles in
ultradegenerate plasmas by a complete evaluation of the on-shell
hard-dense-loop-resummed one-loop fermion self energy for momenta of
the order of the Fermi momentum and above. In the case of zero
temperature, we calculate the nonanalytic terms in the vicinity of
the Fermi surface beyond the known logarithmic approximation, which
turn out to involve fractional higher powers in the energy variable.
For nonzero temperature (but much smaller than the chemical
potential), we obtain the analogous expansion in closed form, which
is then analytic but involves polylogarithms. These
expansions are compared with a full numerical
evaluation of the resulting group velocities and damping coefficients.
\end{abstract}
\maketitle

\section{Introduction}

Unscreened magnetostatic interactions in
a degenerate Fermi gas lead to dramatic changes
of the fermionic dispersion relation in the vicinity
of the Fermi surface. At strictly zero temperature, there
is a logarithmic singularity in the inverse
group velocity, which leads to a breakdown of the
Fermi liquid picture. This effect has been discovered 
in the context of a nonrelativistic degenerate
electron gas by
Holstein, Norton, and Pincus
\cite{Holstein:1973} over thirty years ago, who found
that it gives rise to an anomalous $T \ln T^{-1}$ behavior
of the low-temperature specific heat (see also
\cite{Reizer:1989,Gan:1993,Chakravarty:1995}).

In deconfined degenerate quark matter the same effect
is caused by unscreened chromomagnetic fields,
since the nonperturbative magnetic screening is
parametrically of the order $g^2 T$ and thus
vanishes in the low-temperature limit \cite{Son:1998uk}.
Although chromomagnetic screening may arise in the form
of a Meissner effect in a color superconducting phase,
the appearance of logarithmic terms in the
quark self-energy at (resummed) one-loop order
is also of importance in the case of
color superconductivity, since it leads to a
a significant reduction of the magnitude of
the superconductivity gap 
in a weak-coupling analysis
\cite{Brown:1999aq,
Wang:2001aq}.

In the normal phase of degenerate quark matter, 
non-Fermi-liquid behavior leads to anomalous specific
heat which because of the greater number of gauge bosons
and the stronger coupling is comparatively large.
Moreover, as has been shown in Ref.~\cite{Schafer:2004zf},
the relevant logarithms are stable in the
sense that they do not exponentiate into power-law behavior
when higher loop orders are included
as was previously assumed \cite{Boyanovsky:2000bc,Boyanovsky:2000zj}.
However, an actual numerical evaluation requires to
go beyond the leading logarithmic order calculations
performed in the condensed-matter context 
\cite{Holstein:1973,Reizer:1989,Gan:1993,Chakravarty:1995}.
This has been recently achieved for
the low-temperature specific heat, where
the scale
of the leading temperature logarithm as well as subleading
fractional powers of temperature
were determined in Ref.~\cite{Ipp:2003cj,Gerhold:2004tb},
in a calculation which circumvented a complete
evaluation of the fermion propagator.

Non-Fermi-liquids effects have also been shown recently
to significantly enhance the neutrino emission rate from
normal quark matter \cite{Schafer:2004jp}. However, this calculation
involves the fermionic dispersion relations which have so far been
known only to leading logarithmic accuracy.

In this paper we shall close this gap and
present a complete evaluation of the on-shell
hard-dense-loop-resummed one-loop fermion self energy for momenta of
the order of the Fermi momentum and above, both at zero and
at small temperatures.

\section{Fermion self energy on the light cone}

The fermion self energy is defined through
\begin{equation}
  S^{-1}(P)=S_0^{-1}(P)+\Sigma(P),
\end{equation}
where $S_0(P)=-(P\!\!\!\!/)^{-1}$ is the free fermion propagator,
and $P^\mu=(p^0,\vec p)$ with $p^0=i\omega_n+\mu$,
$\omega_n=(2n+1)\pi T$ in the imaginary time formalism,
and $P^\mu=(E,\vec p)$ after analytic continuation to Minkowski space.
Without loss of generality we shall assume that $\mu>0$.

With the energy projection operators $\Lambda_{\bf p}^\pm={1\over2}(1\pm\gamma_0\gamma^i\hat p^i)$ 
we decompose $\Sigma(P)$ in the quasiparticle and antiquasiparticle self energy,
\begin{equation}
  \Sigma(P)=\gamma_0\Lambda_{\bf p}^+\Sigma_+(P)-\gamma_0\Lambda_{\bf p}^-\Sigma_-(P),
\end{equation}
and
\be
\gamma_0 S^{-1}=\Delta_+^{-1}\Lambda_{\bf p}^+ + 
\Delta_-^{-1}\Lambda_{\bf p}^-
\ee
so that $\Delta_\pm^{-1}=-[p^0\mp(|\mathbf p|+\Sigma_\pm)]$.
 
The one-loop fermion self energy 
is given by
\begin{equation}
  \Sigma(P)=-g^2 C_f T\sum_{\omega}\int {d^3q\over(2\pi)^3}\,\gamma^\mu S_0(P-Q)\gamma^\nu\Delta_{\mu\nu}(Q),
\end{equation}
where $\Delta_{\mu\nu}$ is the 
gauge boson propagator.
Following \cite{Braaten:1991dd} we introduce an intermediate scale $q^*$, such that $m\ll q^*\ll\mu$, 
and we divide the $q$ integration into a soft part ($q<q^*$) and a hard part ($q>q^*$),
\begin{equation}
  \Sigma_+=\Sigma_+^\mathrm{(s)}+\Sigma_+^\mathrm{(h)}.
\end{equation}
For the hard part
we can use the free gluon propagator, whereas for the soft part we have to use a resummed gluon propagator, see below.

The hard contribution to $\Sigma_+$ on the light cone
is most easily computed in covariant Feynman gauge, with the result
\begin{equation}
  \Sigma_+^\mathrm{(h)}={M_\infty^2\over 2p},
\end{equation}
with $M_\infty^2=g^2C_f\mu^2/(4\pi^2)$. 
Here $q^*$ enters only as a correction proportional to $q^*/\mu$, so that
we can send $q^*$ to zero. Correspondingly 
we expect that in the soft contribution
we should be able to send $q^*$ to infinity without encountering
divergences, as will indeed be the case, but only after all
soft contributions are added together.

The leading and next-to-leading contributions to the soft part of the fermion self energy
for momenta of the order of $\mu$ or larger are obtained by
a one-loop diagram where the fermion propagator is a bare propagator,
but the gauge boson propagator is dressed by so-called
hard-dense-loop (HDL) 
\cite{Braaten:1990mz,Altherr:1992mf,Vija:1995is,Manuel:1996td}
self energies, so that
the transverse and longitudinal parts of the gluon propagator 
are given by \cite{LeB:TFT}
\begin{eqnarray}
  \Delta_T(q_0,q)&=&{-1\over q_0^2-q^2-m^2{q_0^2\over q^2}\left[1+{q^2-q_0^2\over2qq_0}
  \log\left({q_0+q\over q_0-q}\right)\right]},\\
  \Delta_L(q_0,q)&=&{-1\over q^2+2m^2\left[1-{q_0\over2q}\log\left({q_0+q\over q_0-q}\right)\right]}.
\end{eqnarray}
We shall consider zero or small temperature $T\ll \mu$, for which
the mass parameter in the HDL propagator is given by
\begin{equation}
  m^2={N_fg^2\mu^2\over4\pi^2},
\end{equation}
which is the asymptotic mass of the transverse modes,
related to the Debye screening mass $m_D$ by $m^2=m_D^2/2$.

On the light cone one finds
the gauge-independent expression
\cite{Manuel:2000mk} 
\begin{eqnarray}
  &&\!\!\!\!\!\!\!\!\!\!\!\!\!\!\!\!
  \Sigma_\pm^{\mathrm{{(s)}}}(E)=-{g^2C_f\over 8\pi^2}\int_0^{{{q^*}}} dq\,q^2\int_{-1}^1 dt\int_{-\infty}^\infty dk_0
  \left[\delta(k_0-k)-\delta(k_0+k)\right]\nonumber\\
  &&\times\Bigg\{2\left(\pm\mathrm{sgn}(k_0)-\hat{\bf p}\!\cdot\!\hat{\bf q}\:\hat{\bf k}\!\cdot\!\hat{\bf q}\right)
  \int_{-\infty}^\infty{dq_0\over2\pi}\rho_T(q_0,q){1+n_b(q_0)-n_f(k_0-\mu)\over k_0+q_0\mp|E|-i\epsilon}\nonumber\\
  &&\quad+\left(\pm\mathrm{sgn}(k_0)+\hat{\bf k}\!\cdot\!\hat{\bf p}\right)
  \bigg[\int_{-\infty}^\infty{dq_0\over2\pi}\rho_L(q_0,q){1+n_b(q_0)-n_f(k_0-\mu)\over k_0+q_0\mp|E|-i\epsilon}\nonumber\\
  &&\quad-{1\over q^2}\left({1\over2}-n_f(k_0-\mu)\right)\bigg]\Bigg\}, \label{r1}
\end{eqnarray}
where ${\bf k}={\bf p}-{\bf q}$ and $E=\pm p$. The distribution functions are given by $n_b(q_0)=1/(e^{q_0/T}-1)$ and
$n_f(k_0-\mu)=1/[e^{(k_0-\mu)/T}+1]$. $\rho_T$ and $\rho_L$ are the spectral densities of tranverse and longitudinal
gauge bosons, respectively,
\be
\rho_{T,L}(q_0,q)=2 {\rm Im} \Delta_{T,L}(q_0+i\epsilon,q).
\ee

We may use $q\ll |E|,k$ because of $q<q^*$ and $|E|\gtrsim\mu$.
Depending on the sign of $E$,
we can drop the term $\delta(k_0+k)$ 
or the term $\delta(k_0-k)$
in Eq. (\ref{r1}), since its contribution is suppressed with $\sim q/E$ 
compared to the remaining contribution. 
Then we find for the soft contribution to the real part of $\Sigma_+$
\begin{eqnarray}
  &&\!\!\!\!\!\!\!\!\!\!
  \real\Sigma_+^\mathrm{(s)}=-{g^2C_f\over 8\pi^2}\int_0^{q^*} dq\,q^2\int_{-1}^1 dt
  \bigg[\int_{-\infty}^\infty{dq_0\over\pi}\left[(1-t^2)\rho_T(q_0,q)+\rho_L(q_0,q)\right]\nonumber\\
  &&\times\:\mathcal{P}{1+n_b(q_0)-n_f(E-\mu-qt)\over q_0-qt}-{1\over q^2}\left(1-2n_f(E-\mu-qt)\right)\bigg].
  \label{si4}
\end{eqnarray}
This quantity vanishes for $E=\mu$ by symmetric integration. After
 performing the $q_0$-integration we therefore have
\begin{eqnarray}
  &&\real\Sigma_+^\mathrm{(s)}=
{g^2C_f\over4\pi^2}\int_0^{q^*} dq\,q^2\int_{-1}^1dt
  \left(n_f(E-\mu-qt)-n_f(-qt)\right)\nonumber\\
  &&\qquad\qquad\qquad\times\left[(1-t^2)\real\Delta_T(qt,q)+\real\Delta_L(qt,q)\right].
\end{eqnarray}
For $\imag\Sigma_+$ (which receives no hard contribution) we find
in an analogous way
\begin{eqnarray}
  &&\!\!\!\!\!\!\!\!\!\!
  \imag\Sigma_+=-{g^2C_f\over 8\pi^2}\int_0^{q^*} dq\,q^2\int_{-1}^1 dt
  \left((1-t^2)\rho_T(qt,q)+\rho_L(qt,q)\right)\nonumber\\
  &&\qquad\times\:\left[1+n_b(qt)-n_f(E-\mu-qt)\right].
\end{eqnarray}

The antiquasiparticle self energy $\Sigma_-^{\mathrm{(s)}}$ is obtained by
inserting negative values of $E$ in the expressions for $\Sigma_+^{\mathrm{(s)}}$ and including an overall factor $(-1)$.
With $\mu>0$ we can
then replace $n_f(E-\mu-qt)$ by 1.

\section{Expansion 
for small $|E-\mu|$ and small $T$}
In this section we will perform an expansion of  $\Sigma_+$ in the region
\begin{equation}
  T\sim |E-\mu|\ll g\mu\ll\mu, \label{ineq}
\end{equation}
where non-Fermi-liquid effects dominate.
We will use the expansion parameter $a:=T/m$,
and we define
$\lambda:=(E-\mu)/T$. From (\ref{ineq}) we have $a\ll1$ and $\lambda\sim\mathcal{O}(1)$. 

In the part with
the transverse gluon propagator we substitute $q=ma^{1/3}z$ and $t=a^{2/3}v/z$. After expanding the
integrand with respect to $a$ we find for the transverse contribution
\begin{eqnarray}
  &&\!\!\!\!\!\!\!\!\!\!
  \real\Sigma^\mathrm{(s)}_{+(T)}=-{g^2C_fma\over\pi^2}\int_{-{q^*\over am}}^{q^*\over am}dv 
  \int_{a^{2/3}|v|}^{q^*\over a^{1/3}m}dz\,{e^\lambda-1\over(1+e^v)(1+e^{\lambda-v})}\nonumber\\
  &&\!\!\!\!\times\bigg[{z^5\over v^2\pi^2+4z^6}+{2v^2z(v^2\pi^2-4z^6)\over(v^2\pi^2+4z^6)^2}a^{2/3}
  -{16v^4z^3(3v^2\pi^2-4z^6)\over(v^2\pi^2+4z^6)^3}a^{4/3}+\ldots\bigg].\nonumber\\ \label{si7}
\end{eqnarray}
The $z$-integrations are straightforward. In the $v$-integrals we may send the integration limits to
$\pm\infty$. Using the formulae
\begin{eqnarray}
  \int_{-\infty}^\infty dv{e^\lambda-1\over(1+e^v)(1+e^{\lambda-v})}|v|^\alpha
  &=&\Gamma(\alpha+1)\left[\Li_{\alpha+1}(-e^{-\lambda})-\Li_{\alpha+1}(-e^{\lambda})\right]
  \quad\forall \alpha\ge0,\quad\\
  \int_{-\infty}^\infty dv{e^\lambda-1\over(1+e^v)(1+e^{\lambda-v})}\log|v|
  &=&-\gamma_E\lambda+{\partial\over\partial \alpha}
  \left(\Li_{\alpha+1}(-e^{-\lambda})-\Li_{\alpha+1}(-e^{\lambda})\right)\Big|_{\alpha=0},
\end{eqnarray}
we find, neglecting terms which are suppressed at least with $(m/q^*)^4$,
\begin{eqnarray}
  &&\!\!\!\!\!\!\!\!\!\!\!\! \real\Sigma^\mathrm{(s)}_{+(T)}=-g^2C_fm\nonumber\\
  &&\!\!\!\!\!\times\Bigg\{{a\over12\pi^2}\left[\lambda\log\left({2(q^*)^3\over a m^3\pi}\right)+\gamma_E\lambda
  -{\partial\over\partial \alpha}\left(\Li_{\alpha+1}(-e^{-\lambda})-\Li_{\alpha+1}(-e^{\lambda})\right)\Big|_{\alpha=0}
  \right]\nonumber\\
  &&+{2^{1/3}a^{5/3}\over9\sqrt{3}\pi^{7/3}}\Gamma\left(\textstyle{5\over3}\right)
  \left(\Li_{5/3}(-e^{-\lambda})-\Li_{5/3}(-e^\lambda)\right)\nonumber\\
  &&-20{2^{2/3}a^{7/3}\over27\sqrt{3}\pi^{11/3}}\Gamma\left(\textstyle{7\over3}\right)
  \left(\Li_{7/3}(-e^{-\lambda})-\Li_{7/3}(-e^\lambda)\right)\nonumber\\
  &&+{8(24-\pi^2)a^3\log a\over27\pi^6}\lambda(\lambda^2+\pi^2)+\mathcal{O}(a^3)\Bigg\}.
\end{eqnarray}

In the longitudinal part we substitute $q=mx$ and $t=au/x$. In  a similar way as for the transverse part
we find
\begin{eqnarray}
  \real\Sigma^\mathrm{(s)}_{+(L)}=-g^2C_fm\left[ {a\lambda\over8\pi^2}\log\left({2m^2\over(q^*)^2}\right)
  -{(\pi^2-4)a^3\log a\over96\pi^2}\lambda(\lambda^2+\pi^2)+\mathcal{O}(a^3) \right].
\end{eqnarray}

Turning now to $\imag\Sigma_+$ we notice that it vanishes at $E=\mu$
only in the case of $T=0$. For finite temperature, however small,
there is an IR divergent contribution in the transverse sector
\cite{LeB:TFT},
\be
\imag\Sigma_{+(T)}^\mathrm{(s)}\big|_{E=\mu}=
-{g^2 C_f T\04\pi} \ln{m\0\Lambda_{\rm IR}}
\ee
where the infrared cutoff may be provided at finite temperature
by the nonperturbative magnetic screening mass of QCD. In QED,
where no magnetostatic screening is possible, a resummation
of these singularities leads to nonexponential damping behavior
\cite{Blaizot:1996az}.

After subtraction of the energy independent part we have
\begin{eqnarray}
  &&\imag\Sigma_+^\mathrm{(s)}
-\imag\Sigma_+^\mathrm{(s)}\big|_{E=\mu}
={g^2C_f\over8\pi^2}\int_0^{q^*} dq\,q^2\int_{-1}^1dt
  \left(n_f(E-\mu-qt)-n_f(-qt)\right)\nonumber\\
  &&\qquad\qquad\qquad\qquad\qquad\qquad
\times\left[(1-t^2)\rho_T(qt,q)+\rho_L(qt,q)\right].
\end{eqnarray}
Following the
steps which led to Eq. (\ref{si7}), we find for the transverse contribution
\begin{eqnarray}
  &&\!\!\!\!\!\!\!\!\!\!\!\!\!\!\!\!\!\!\!\!
  \imag\Sigma_{+(T)}^\mathrm{(s)}-\imag\Sigma_{+(T)}^\mathrm{(s)}\big|_{E=\mu}
  ={g^2C_fma\over2\pi}\int_{-{q^*\over am}}^{q^*\over am}dv 
  \int_{a^{2/3}|v|}^{q^*\over a^{1/3}m}dz\,{e^\lambda-1\over(1+e^v)(1+e^{\lambda-v})}\nonumber\\
  &&\qquad\times\bigg[-{z^2v\over v^2\pi^2+4z^6}+{16v^3z^4\over(v^2\pi^2+4z^6)^2}a^{2/3}
  +{16v^5(v^2\pi^2-12z^6)\over(v^2\pi^2+4z^6)^3}a^{4/3}+\ldots\bigg].\nonumber\\
\end{eqnarray}
Using the formula
\begin{eqnarray}
  &&\!\!\!\!\!\!\!\!\!
  \int_{-\infty}^\infty dv{e^\lambda-1\over(1+e^v)(1+e^{\lambda-v})}
  |v|^\alpha\mathrm{sgn}(\alpha)\nonumber\\
  &&\!\!\!\!=-\Gamma(\alpha+1)\left[\Li_{\alpha+1}(-e^{-\lambda})+\Li_{\alpha+1}(-e^{\lambda})
  +2\left(1-2^{-\alpha}\right)\zeta(\alpha+1)\right]
  \quad\forall \alpha\ge0 \nonumber\\
\end{eqnarray}
we find in a similar way as above
\begin{eqnarray}
  &&\!\!\!\!\!\!\!\!\!\!\!\!\!\!\!\!\!\!\!\!\!\!\!\!\!\!\!\!\!\!\!\!
  \imag\Sigma_{+(T)}^\mathrm{(s)}-\imag\Sigma_{+(T)}^\mathrm{(s)}\big|_{E=\mu}
  ={g^2C_fm}\biggl\{-{a\over12\pi}\log\cosh\left({\lambda\over2}\right)\nonumber\\
  &&-{2^{1/3}a^{5/3}\over9\pi^{7/3}}\Gamma\left(\textstyle{5\over3}\right)
  \left[\Li_{5/3}(-e^{-\lambda})+\Li_{5/3}(-e^\lambda)+2\left(1-2^{-2/3}\right)
  \zeta\left(\textstyle{5\over3}\right)\right]\nonumber\\
  &&-20{2^{2/3}a^{7/3}\over27\pi^{11/3}}\Gamma\left(\textstyle{7\over3}\right)
  \left[\Li_{7/3}(-e^{-\lambda})+\Li_{7/3}(-e^\lambda)+2\left(1-2^{-4/3}\right)
  \zeta\left(\textstyle{7\over3}\right)\right]+\mathcal{O}(a^3)\biggr\}.
\end{eqnarray}
For the longitudinal part we obtain
\begin{equation}
  \imag\Sigma_{+(L)}^\mathrm{(s)}-\imag\Sigma_{+(L)}^\mathrm{(s)}\big|_{E=\mu}
  =-{g^2C_fm}\left[{a^2\lambda^2\over64\sqrt{2}}+\mathcal{O}(a^3)\right].
\end{equation}

We remark that the determination of the coefficient of the $\mathcal{O}(a^3)$ terms in $\Sigma_+$ would 
require resummation of IR enhanced contributions along the lines of Ref. \cite{Gerhold:2004tb}, App. A.

Putting the pieces together, and using the abbreviation $\emu=E-\mu$, 
we obtain for the real part
\begin{eqnarray}
  &&\!\!\!\!\!\!\!\!\!\!\!\!\!\!\!
  \real\Sigma_+={M_\infty^2\over 2E} - g^2C_fm\, \mathrm{sgn}(\emu)
  \Bigg\{{|\emu|\over12\pi^2m}\left[\log\left({4\sqrt{2}m\over\pi T f_1(\emu/T)}\right)+1\right]
  +{2^{1/3}\sqrt{3}\over45\pi^{7/3}}\left({T\over m}f_2\left({\emu\over T}\right)\right)^{5/3}\nonumber\\
  &&-20{2^{2/3}\sqrt{3}\over189\pi^{11/3}}\left({T\over m}f_3\left({\emu\over T}\right)\right)^{7/3}
  -{6144-256\pi^2+36\pi^4-9\pi^6\over864\pi^6}\left({T\over m} 
  f_4\left({\emu\over T}\right)\right)^3\log\left({m\over T}\right)\nonumber\\
  &&+\mathcal{O}\bigg(\left({T\over m}\right)^3\bigg)\Bigg\},
\end{eqnarray}
where
\begin{eqnarray}
  f_1(\lambda)&=& \exp\left[1-\gamma_E+{1\over\lambda}{\partial\over\partial\alpha} 
  \left(\Li_{\alpha+1}(-e^{-\lambda})-\Li_{\alpha+1}(-e^{\lambda})\right)\Big|_{\alpha=0}\right],\\
  f_2(\lambda)&=& \left|\Gamma\left(\textstyle{8\over3}\right)
  \left(\Li_{5/3}(-e^{-\lambda})-\Li_{5/3}(-e^{\lambda})\right)\right|^{3/5} ,\\
  f_3(\lambda)&=& \left|\Gamma\left(\textstyle{10\over3}\right)
  \left(\Li_{7/3}(-e^{-\lambda})-\Li_{7/3}(-e^{\lambda})\right)\right|^{3/7} ,\\
  f_4(\lambda)&=& \left|\lambda(\lambda^2+\pi^2)\right|^{1/3}.
\end{eqnarray}
We note that the dependence on $q^*$ indeed
drops out in the sum of the transverse and longitudinal parts.

In the zero temperature limit ($|\lambda|\to\infty$) we have $f_i(\lambda)\to|\lambda|$. If the temperature
is much higher than $|E-\mu|$ (i.e. $\lambda\to0$) we have $f_1(\lambda)\to c_0:={{\pi\over2}\exp(1-\gamma_E)}=2.397357\ldots$ 
and $f_{2,3,4}(\lambda)\to 0$.
For $|\lambda|\gg c_0$ or $|\lambda|\ll c_0$ we may approximate $f_1(\lambda)$ with $\mathrm{max}(c_0,|\lambda|)$,
which is qualitatively the result quoted in \cite{Brown:1999aq}. 
It should be noted, however, that the calculation of
Ref. \cite{Brown:1999aq} only took into account transverse gauge bosons, 
and therefore the scale under the logarithm
and its parametric dependence on the coupling was not correctly rendered.

For the imaginary part we find
\begin{eqnarray}
  && \imag\Sigma_{+}-\imag\Sigma_{+}\big|_{E=\mu}
  =g^2C_fm\bigg[-{T\over24\pi m}g_1\left({\emu\over T}\right)
  +3{2^{1/3}\over45\pi^{7/3}}\left({T\over m}g_2\left({\emu\over T}\right)\right)^{5/3}\nonumber\\
  &&\quad\qquad\qquad
- {1\over64 \sqrt{2}}\left({T\over m}g_3\left({\emu\over T}\right)\right)^{2}
  +20{2^{2/3}\over63\pi^{11/3}}\left({T\over m}g_4\left({\emu\over T}\right)\right)^{7/3}
  +\mathcal{O}\bigg(\left({T\over m}\right)^3\bigg)\bigg],
\end{eqnarray}
where
\begin{eqnarray}
  g_1(\lambda)&=&2\log\cosh\left({\lambda\over2}\right),\\
  g_2(\lambda)&=&\left[-\Gamma\left(\textstyle{8\over3}\right)
  \left(\Li_{5/3}(-e^{-\lambda})+\Li_{5/3}(-e^\lambda)+2\left(1-2^{-2/3}\right)
  \zeta\left(\textstyle{5\over3}\right)\right)\right]^{3/5},\\
  g_3(\lambda)&=&|\lambda|,\\
  g_4(\lambda)&=&\left[-\Gamma\left(\textstyle{10\over3}\right)
  \left(\Li_{7/3}(-e^{-\lambda})+\Li_{7/3}(-e^\lambda)+2\left(1-2^{-4/3}\right)
  \zeta\left(\textstyle{7\over3}\right)\right)\right]^{3/7}.
\end{eqnarray}
In the zero temperature limit we have $g_i(\lambda)\to|\lambda|$. If the temperature
is much higher than $|E-\mu|$  we have $g_i(\lambda)\to0$.

Explicitly, our $T=0$ result reads
\begin{eqnarray}
  \Sigma_+\big|_{T=0}&=&{M_\infty^2\over 2E}-g^2C_fm\,
  \Bigg\{{\emu\over12\pi^2m}\left[\log\left({4\sqrt{2}m\over\pi|\emu|}\right)+1\right]+{i|\emu|\over24\pi m}\nonumber\\
  &&\qquad+{2^{1/3}\sqrt{3}\over45\pi^{7/3}}\left({|\emu|\over m}\right)^{5/3}(\mathrm{sgn}(\emu)-\sqrt{3}i)\nonumber\\&&\qquad
  + {i\over64 \sqrt{2}}\left({\emu\over m}\right)^2
  -20{2^{2/3}\sqrt{3}\over189\pi^{11/3}}\left({|\emu|\over m}\right)^{7/3}(\mathrm{sgn}(\emu)+\sqrt{3}i)\nonumber\\
  &&\qquad-{6144-256\pi^2+36\pi^4-9\pi^6\over864\pi^6}\left({\emu\over m} 
  \right)^3 \left[\log\left({{0.928}\,m\over |\emu|}\right) 
-{i\pi\mathrm{sgn}(\emu)\over 2}  \right]\nonumber\\&&
  \qquad+\mathcal{O}\bigg(\left({|\emu|\over m}\right)^{11/3}\bigg) \Bigg\},
\end{eqnarray}
{where the scale of the last logarithm was determined by resumming IR enhanced contributions \cite{andreasdiss}.}

Apart from the first logarithmic term, the leading imaginary parts
contributed by the transverse and longitudinal gauge bosons were
known previously \cite{LeBellac:1996kr,Vanderheyden:1996bw,Manuel:2000mk}.
As our results show,
the damping rate obtained by adding these two leading terms 
\cite{LeBellac:1996kr,Manuel:2000mk} is actually incomplete 
beyond the leading term, 
because the subleading transverse term of order $|\emu|^{5/3}$
is larger than the leading contribution from $\Sigma_{L}$.

\section{Numerical results and discussion}

At large values of $E-\mu$ or at large negative values of $E$,
where one obtains the self energy of the antiquasiparticles,
the soft contribution ${\real}\Sigma^s$ can be shown to vanish.
At $T=0$, where the imaginary part does not contain
an infrared divergent contribution, $\imag\Sigma_{+}$
approaches the constant\footnote{The numerical constant in Eq.~(\ref{asimagS}) agrees
with the value given with two significant digits
in Ref.~\cite{Vanderheyden:1996bw} (taking into account the
different normalization), and it is
very close to, but not in complete agreement with, the value
$({1\024\pi}+{1\064})\sqrt2 \approx 0.040854$ quoted in
Ref.~\cite{Manuel:2000mk}.}
\be\label{asimagS}
\lim_{E\to\infty}
\imag\Sigma_{+}(E)\Big|_{T=0}=-g^2 C_f m \times {0.040534\ldots} 
\ee
The resulting damping constant $\gamma=-\imag\Sigma_{+}$
is also that of the antiquasiparticles, which are of course
far from their nonexistent Fermi surface for $\mu>0$.

\begin{figure}
\centerline{\includegraphics[scale=0.65
]{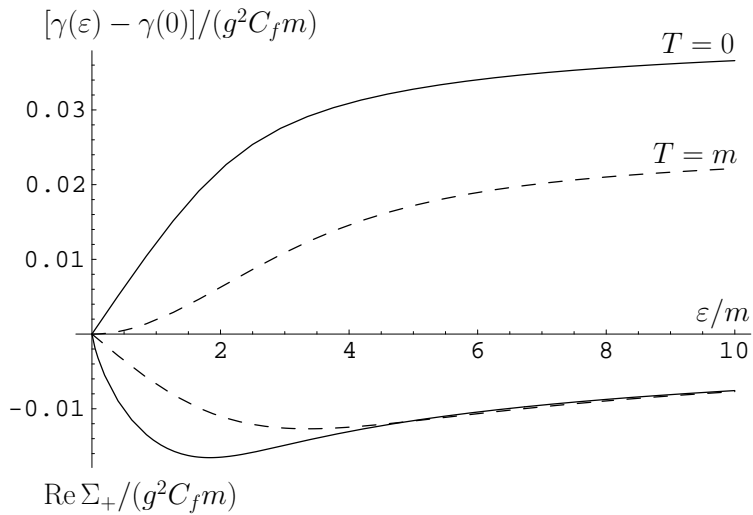}}
\caption{Real and imaginary part of 
$\Sigma^{(s)}_+/(g^2C_f m)$ as a function of $\varepsilon/m \equiv {(E-\mu)/m}$
at $T=0$ (full lines) and $T=m$ (dashed lines).
\label{freimsigma}}
\end{figure}

At intermediate energies $|E-\mu| \gtrsim m$, both the real and
imaginary parts of $\Sigma_+$ are nontrivial functions that
we have evaluated numerically. The results are shown in Fig.~\ref{freimsigma}
for the two cases $T=0$ and $T=m$. Since at finite temperature
the imaginary part of $\Sigma$ contains a (constant) infrared
singular contribution, we plot $\gamma(E-\mu)-\gamma(0)$ instead.
The latter function is even with respect to its argument, resulting
in a cusp at $E=\mu$ for $T=0$, while at finite $T$ the
damping contribution vanishes quadratically at $E=\mu$.
The real part is an odd function with respect to $E-\mu$.
Again, $E=\mu$ is a nonanalytic point at $T=0$, but analytic
at finite $T$.

\begin{figure}
\centerline{\includegraphics[scale=0.65
]{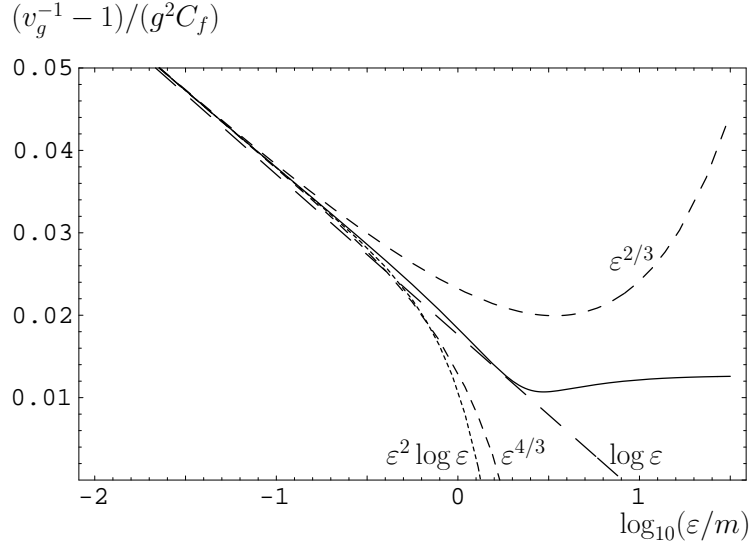}}
\caption{$(v_g^{-1}-1)/(g^2C_f)$ as a function of $\log_{10}((E-\mu)/m)$
at zero temperature. 
\label{finvv}}
\end{figure}

At $T=0$, the consequence of the nonanalyticity at $E=\mu$
is that the group velocity $dE/dp$ together with the
residue in the propagator vanishes.
The group velocity is determined by
\be
v_g^{-1}=1+{M_\infty^2\0 2E^2}-{\partial \real \Sigma_+^{\mathrm{{(s)}}}\0 \partial E}.
\ee
The numerical result for $v_g^{-1}-1$ is given in
Fig.~\ref{finvv} together with the series expansion
for small $|E-\mu|$ following from the results of the previous section.
As one can see, this expansion converges well only for $|E-\mu| \ll m$,
and it turns out that the logarithmic contribution
\be
v_g^{-1}(E-\mu)=1+{g^2 C_f\012\pi^2} \left[ \ln{4 \sqrt2\, m\0\pi|E-\mu|}+{3\02}
\right]+O\bigg(\left({E-\mu\0m}\right)^{2/3}\bigg)
\ee
is already a rather good approximation up to the point where
one should switch to the leading order result
$M_\infty^2/( 2E^2) =  g^2 C_f/(8\pi^2) \approx 0.013g^2 C_f$ that is relevant for
larger values of $m \lesssim| E-\mu| \ll \mu$.

At small finite temperature the results of the previous section
show that
the growth of $v^{-1}_g$ for
$E\to\mu$ is limited by
\be
v_g^{-1}=1+{g^2 C_f\012\pi^2} \ln{c_0' m\0T} 
+ O\bigg(\left({T\0m}\right)^3\bigg)
\ee
with $c_0'=4\sqrt{2}e^{5/2}/(\pi c_0)\approx 9.15016$.
In this case
the group velocity at and above the Fermi surface
can be approximated by
\be\label{vgapprox}
v_g^{-1}(E-\mu)\approx
1+{g^2 C_f\012\pi^2}\, {\rm max} \left\{ {\rm min} \left(
\ln{9.15\, m\0T}, \ln{8.07\, m\0|E-\mu|}\right)  ,{3\02} \right\}.
\ee

This result for the group velocity of the quasiparticle excitations
is for example
of direct relevance for the calculation of the neutrino
emission from normal degenerate quark matter
\cite{Iwamoto:1980eb} which is enhanced by
non-Fermi-liquid effects \cite{Schafer:2004jp}.
As was shown recently in Ref.~\cite{Schafer:2004jp},
the neutrino emissitivity involves two powers of $\alpha_s\ln(m/T)$
from the quasiparticle group velocities, which overcompensate
the single power of $\alpha_s\ln(m/T)$ in the specific heat
that counteracts in the cooling rate. For a numerical evaluation
one evidently needs to know the constants under these logarithms.
Interestingly enough, the constants under the log's
in (\ref{vgapprox}) are much larger than the constant under
the log arising in the specific heat
determined in Ref.~\cite{Ipp:2003cj,Gerhold:2004tb} as $\log(0.282m/T)$ ---
for more discussion see Ref.~\cite{andreasdiss}.

\section{Conclusions}
In this article we have computed the fermion self energy in an ultradegenerate relativistic plasma.
For small $|E-\mu|$ and small $T$ we have obtained a perturbative expansion of $\Sigma_+$ beyond the leading logarithm that is responsible
for non-Fermi-liquid behavior. We found that dynamical screening
leads to fractional powers in this series, which are analogous to the fractional powers in the anomalous specific 
heat in normal degenerate quark matter \cite{Ipp:2003cj,Gerhold:2004tb}.
Furthermore we have performed a numerical computation of the self energy and the group velocity for larger values 
of $|E-\mu|$.
Our results provide an important ingredient for quantitative calculations
of non-Fermi-liquid effects such as
the computation of the enhanced neutrino emissivity of ungapped quark
matter 
\cite{Schafer:2004jp}.

\acknowledgments
This work has been 
supported by the Austrian Science Foundation FWF, project no. P16387-N08.


\end{document}